\documentclass[12pt,preprint]{aastex}

\slugcomment{Accepted by ApJ (scheduled publication in 
volume 594, September 1, 2003)}

\shorttitle{}
\shortauthors{Martin et al.}

\begin{document}


\title{An HST/WFPC2 
Survey for Brown Dwarf Binaries in the $\alpha$Persei 
and the Pleiades Open Clusters}


\author{Eduardo L. Mart\'\i n}
\affil{Institute for Astronomy, University of Hawaii at Manoa, 
2680 Woodlawn Drive, Honolulu HI 96822, USA}
\email{ege@ifa.hawaii.edu}

\author{David Barrado y Navascu\'es}
\affil{LAEFF-INTA, Apdo. 50727, 28080 Madrid, Spain}
\email{barrado@laeff.esa.es}

\author{Isabelle Baraffe}
\affil{UMR 5574 CNRS, Ecole Normale Sup\'erieure, 69364 Lyon Cedex 07, 
France}
\email{ibaraffe@ens-lyon.fr}

\author{Herv\'e Bouy}
\affil{E.S.O, Karl Schwarzschildstra\ss e 2, D-85748 Garching, Germany}
\email{hbouy@eso.org}

\author{Scott Dahm}
\affil{Institute for Astronomy, University of Hawaii at Manoa, 
2680 Woodlawn Drive, Honolulu HI 96822}
\email{dahm@ifa.hawaii.edu}




\newpage

\begin{abstract}
 
We present the results of a high-resolution imaging survey for brown dwarf 
binaries in two open clusters. The observations were  
carried out with the Wide Field Planetary Camera~2 onboard the 
Hubble Space Telescope. 
Our sample consists of 8 brown dwarf candidates in $\alpha$Persei and 
25 brown dwarf candidates in the Pleiades. We have resolved 4 binaries in 
the Pleiades 
with separations in the range 0$\arcsec$.094--0$\arcsec$.058, 
corresponding to projected separations between 11.7~AU and 7.2~AU. 
No binaries were found among the $\alpha$Persei targets. 
Three of the binaries have proper motions consistent with cluster membership 
in the Pleiades cluster, 
and for one of them we report the detection of H$_\alpha$ in emission and 
LiI absorption obtained from Keck~II/ESI spectroscopy.  
One of the binaries does not have a proper motion consistent with Pleiades 
membership. 
We estimate that brown dwarf binaries wider than 12~AU are less frequent than 
9\% in the $\alpha$Persei and Pleiades clusters. This is consistent with an 
extension to substellar masses of a trend observed among stellar binaries: the maximum 
semimajor axis of binary systems decreases with decreasing primary mass. 
We find a binary frequency of 2 binaries over 13  brown dwarfs with confirmed 
proper motion membership in the Pleiades, corresponding to a binary fraction of 
15$^{+15}_{-5}$\% . These binaries are limited to the separation range 
7-12~AU and their mass ratios are larger than 0.7. The observed properties 
of Pleiades brown dwarf binaries appear to be similar to their older  
counterparts in the solar neighborhood. The relatively high binary frequency 
($\ge$ 10\%), the bias to separations smaller than about 15 AU and the trend 
to high mass ratios (q$\ge$0.7) 
are fundamental properties of brown dwarfs. Current theories of brown 
dwarf formation do not appear to provide a good description of all these properties.    
  
\end{abstract}


\keywords{stars: low mass, brown dwarfs, binaries: visual,  
techniques: high angular resolution, surveys}


\section{Introduction}

Brown dwarfs (BDs) are very low-mass objects that do not stabilize 
on the hydrogen-burning main sequence because they develope degenerate 
cores. 
They are akin to giant planets in that 
their luminosity and temperature drop continuously with time, 
and ultimately they become extremely cool and faint. 
The borderline between stars and BDs has been known for 40 years to be 
at about 0.08~solar mass (M$_\odot$) 
for solar metallicity  (Hayashi \& Nakano 1963; Kumar 1963). 
Recent calculations yield a substellar limit at 0.072~M$_\odot$ 
(e.g. Baraffe et al. 1998).  

For many years BDs eluded firm detection, but since 1995  
(Nakajima et al. 1995; Rebolo, Zapatero Osorio, \& Mart\'\i n 1995) 
there has been a pletora of discoveries. 
The evidence for BDs is based on observations of lithium, 
luminosity and surface temperatures of faint dwarfs, as well as 
astrometric and radial velocities of main sequence stars. 
Dynamical measurements of BD 
masses are becoming available with the study of nearby 
binaries where both components are substellar 
(Mart\'\i n et al. 2000b; Lane et al. 2001). 

Recent censuses of BDs in young open clusters 
indicate that they are numerous 
(B\'ejar et al. 2001; Moraux et al. 2003). 
With increasing age, BDs cool 
down and reach very faint luminosities, becoming indeed 
very difficult to detect. Young stellar associations and open clusters 
have provided a gold mine for BD searches because these objects 
are still in the early phase of gravitational contraction. 
Hence, they are brighter and warmer than older BDs. 

Several processes of BD formation have been studied in some detail 
in the last few years. 
They can be divided in two main groups: (1) Failed star-formation model: 
BDs are ejected from unstable multiple systems during the collapse and 
fragmentation of a turbulent molecular cloud (Reipurth \& Clarke 2001, 2003; 
Sterzik \& Durisen 2003); 
(2) Small star-formation model: BDs result from the collapse of small cores of 
substellar mass in supersonically turbulent molecular clouds 
(Padoan \& Nordlung 2002). 

These processes lead to different outcomes in the overall properties of a 
brown dwarf population which result in observable effects. 
Hydrodynamical simulations show that the failed star-formation model produces 
a very low frequency ($<$5\%) of BD binaries, 
which must be limited to systems closer than 10~AU 
(Bate et al. 2002; Durisen, Sterzik \& Pickett 2001; 
Delgado-Donate \& Clarke 2003). 
The consequences of the small star-formation model for the properties of the 
BD binary population have not been explored, 
but it is reasonable to expect a higher frequency of binaries 
than in the previous scenario.  
  
Using the high-resolution capability of the Hubble Space Telescope (HST), 
Keck, and Adaptive Optics (AO) on large ground-based telescopes, the first 
very low-mass binaries have been resolved 
(Mart\'\i n et al. 1998a, 1999, 2000a,b; Bouy et al. 2003; 
Burgasser et al. 2003; 
Close et al. 2002, 2003; Gizis et al. 2003; Koerner et al. 1999; 
Reid et al. 2001). 
Most of the work has concentrated on nearby field ultracool dwarfs. 
However, the field targets represent a mix of ages, distances, 
primary masses and metallicities. 
Hence, the results are difficult to interpret. 
In contrast, an open cluster sample has well-known age, distance, 
primary masses and metallicity. 
In this paper we present the first confirmed cluster BD binaries 
resolved with HST.  

The paper is organized as follows: 
In Section 2 we describe the target selection and observations. 
In Section~3 we present the photometric and point-spread-function (PSF) 
analysis. 
Section~4 contains the discussion of our results, and in Section~5 we 
present our final remarks and suggestions for future work.

\section{Target Selection and Observations}

The targets were compiled from several lists of very low-mass 
(VLM, M$<$ 0.1 M$_{\odot}$) candidate members. 
For the Pleiades we used the information provided in the 
works of Mart\'\i n et al. (1996, 1998b), 
Rebolo et al. (1996) and Zapatero Osorio et al. (1997) for 
Calar objects;  Zapatero Osorio et al. (1999) for  
Roque objects; Bouvier et al. (1998), Stauffer, Schultz \& Kirkpatrick (1998) 
and Mart\'\i n et al. (2000a) for CFHT-PL objects;  
Hambly et al. (1999) for IPMBD objects; and Festin (1998a,b) for NPL objects.  
These compilations include most of the 
known Pleiades VLM stars and brown dwarf candidates (BDCs) that had not been 
previously observed with HST. 
Of the 29 objects originally proposed for this program, 26 have been observed.
One of them turned out to be a duplication because NPL~40 is the same 
object as Roque~33. This object was observed twice. 
The remaining 3 Pleiades targets were not scheduled for observation. 
For $\alpha$Per we used the data in Rebolo, Mart\'\i n \& Magazz\`u (1992), 
Prosser (1994), Basri \& Mart\'\i n (1998),  and 
Stauffer et al. (1999). We selected 14 VLM stars and BDCs, and 8 
were observed. The other 6 were withdrawn. 

Observations were carried out between July 2000 and August 2001 
as part of an HST Snapshot (SNAP-8701, HST Cycle~9) program
designed to fill short intervals between accepted GO observations.
Each BDC was centered in the PC1 chip of the 
Wide-Field Planetary Camera~2 (WFPC2).  With a plate-scale of 0$\arcsec$.0455 
per pixel, the PC1 has a field of view of 36$\arcsec$ in diameter.
Assuming a distance of 125 pc for the Pleiades, this provides a
physical separation of 2250~AU from each BDC to search
for companions.  The observations were made in
two broadband filters, the F814W and the F785LP with central wavelengths
at 792.4~nm and 862.1~nm respectively.  The filters were chosen
to provide high throughput and a clear color separation between
BDs and background stars and galaxies.
Two exposures were taken in each filter to allow for proper cosmic
ray rejection, yielding total integration times of 600~s and 280~s 
 in the F814W and F785LP filters, respectively.

The binarity of IPMBD~25 was noticed in an early analysis of a subset of our 
WFPC2 dataset reported by Dahm \& Mart\'\i n (2003). In order to confirm the 
cluster membership, we decided to apply the lithium test (Magazz\`u, 
Mart\'\i n \& Rebolo 1993). 
We observed IPMBD~25 on 15 November 2001 with the Echellette Spectrograph 
and Imager (ESI, 
Sheinis et al. 2002) 
mounted on the Cassegrain focus of the Keck~II telescope. 
We used the echellette 
mode to cover the entire spectrum from 0.39 to 1.1~$m$m. 
The spectral resolution 
given by the 0$\arcsec$.6 slit was 6500. 
Three exposures of 1200~s 
were obtained. Data reduction was made using IRAF's \emph{echelle} tasks. 
The spectra were bias subtracted, flat fielded, and combined. 
Wavelength calibration was made using a CuAr lamp spectrum obtained the 
same night.


\section{Data Analysis and Products}

\subsection{HST/WFPC2 data} 

We used the data reduced by the HST pipeline. 
Aperture photometry was accomplished
using the IRAF\footnote{IRAF is distributed by National Optical Astronomy 
Observatories, 
which is operated by the Association of Universities for Research in 
Astronomy, Inc., 
under contract with the National Science Fundation.} \emph{phot} task in 
\emph{daophot}.
Photometry was completed for all point sources in the PC1 chip using the 
recommended aperture of 11 pixels, corrected for finite aperture 
(subtract 0.1 mag). 
Magnitudes are given in Table~1 in the VEGA system using the zeropoints 
provided in the WFPC2 User Manual (F814W=21.639, F785LP=20.688). 
Roque~33 was observed twice because it was also entered in the target list 
as NPL~40. 
We report the average values of the photometry for both observations. 
Our photometric measurements are in good agreement with those reported in 
the literature, 
except for those given by Zapatero Osorio et al. (1999) which are 
systematically brighter. 
Jameson et al. (2002) noted that the Zapatero Osorio et al.'s $I$-band 
data lies on the 
Harris system, and they provide a conversion to the Cousins system. 
Our F814W data is a good approximation to the $I$-band Cousins system. 
We estimate that the 5~$\sigma$ limiting magnitude for the survey is
approximately I$_{C}\sim$22.5 and was constrained by the lower throughput
and shorter exposure times of the F785LP images.   

All images have been inspected visually to search for obviously resolved 
companions. 
There were not any companion BDCs in the PC1 chip for separations larger than 
3 pixels, i.e. 0$\arcsec$.13 for a nominal scale of 
0$\arcsec$.0455~pix$^{-1}$. 
We first identified the binary candidates as objects with larger FWHM than 
average using the IRAF task \emph{imexam}. 
Figure~1 illustrates the results of the FWHM analysis for the 
F814W filter. The results for the F785LP filter were very similar.  
We identified 4 binary candidates with FWHM$>$2 pixels in both the F814W 
and F785LP filters, 
namely CFHT-Pl-12 and 19, IPMBD~25 and 29. 

We used {\bf a custom-made} PSF fitting program to compute the precise separation, 
position angle and flux ratio of each candidate binary system. 
This program is identical to that used by Bouy et al. (2003), and it is fully described 
in that paper. A brief summary is given here for compleness: 
The PSF fitting routine builds a model PSF using 8 different PSF stars in each filter 
coming from different WFPC2 images. 
A cross correlation between the model and the binary system yields  
the best values for five free parameters (flux ratio and the pixel coordinates of the two 
components of the binary system). 
By using non-linear PSF fitting we were able to push the limit 
of detection down to $\sim$ 0$\arcsec$.060 arcsec for non-equal luminosity systems. 
Figure~2 displays one example of the PSF fitting technique. 
The uncertainties and limitations of this technique are discussed in Bouy et al. (2003). 
The final best solutions for the binary parameters are summarized in Table~2. 
 
\subsection{Keck~II data} 

We report the detection of H$_\alpha$ in emission (equivalent width = 13.1 $\pm$0.2 $\AA$), 
and LiI resonance line in absorption (equivalent width = 0.55 $\pm$0.10 $\AA$) in the 
ESI spectrum of IPMBD~25 (shown in Figure~3). 
The lithium depletion boundary (LPD) in the Pleiades is located at $I$=17.8$\pm$0.1 
(Stauffer et al. 1998). 
Cluster members fainter than the LPD are expected to have preserved a measurable amount of lithium 
in their photosphere. Cluster members brighter than the LPD are expected to have depleted their 
lithium abundances to a level that cannot be detected. 
The apparent $I$ magnitude of IPMBD~25 is 17.67 (Hambly et al. 1999), which is brighter 
than the lithium boundary. However, the apparent magnitude of the primary is $I$=17.93$\pm$0.09, 
which is fainter than the boundary. Thus, our lithium detection confirms the 
membership of IPMBD~25 in the cluster, and is consistent with the previously established LPD. 
The LiI equivalent width that we measured in our ESI spectrum of IPMBD~25 is very similar 
to Stauffer et al.'s measurement for CFHT-Pl-11.

\section{Discussion} 

\subsection{Cluster membership and binary frequency} 

Two proper motion studies of 
VLM stars and BDs in the Pleiades are available 
(Hambly et al. 1999; Moreaux et al. 
2001). 14 of our targets have confirmed proper motion membership. 
10 do not have proper motion information available. 
2 do not have proper motions consistent 
with cluster membership. The proper motion nonmembers are: CFHT-Pl-15 and CFHT-Pl-19. 
The first one has 
lithium (Stauffer et al. 1998) but we do not consider the Li detection to be sufficient to rank it 
as a {\it bona fide} cluster member. Young BDs are not uncommon in the general direction 
of the Pleiades 
cluster. CFHT-Pl-15 may be one more example of a scattered population of young BDs in the 
Taurus-Auriga region (Oppenheimer et al. 1997). 
CFHT-Pl-19 is a binary. It is unlikely that the binarity has affected the proper motion 
measurement of 
Moreaux et al. (2001). Furthermore, its position in the H-R diagram is not consistent with 
being a cluster binary (Mart\'\i n et al. 2000a). 

We consider CFHT-Pl-12, IPMBD~25 and 29 as {\it bona fide} Pleiades binaries. All of them have 
proper motion membership. CFHT-Pl-12 and IPMBD~25 have lithium detections. The lithium test has 
not been applied in IPMBD~29. 

IPMBD~25 and 29 were found in a shallower and wider survey than CFHT-Pl-12. 
IPMBD~25~A has $I$=17.93$\pm$0.09 which is brighter than the sensitivity limit of Hambly et al.'s 
survey 
($I$=18.4). However, IPMBD~29~A has $I$=18.70$\pm$0.15 which is fainter than the survey limit.
IPMBD~29 was found by Hambly et al. (1999) because it is a binary. Thus, we must exclude it from 
our binary frequency statistics. 

Our sample of {\it bona fide} cluster members includes 13 objects, which include 2 resolved binaries. 
Thus, we derive a binary frequency (defined as number of binaries divided by total number of objects 
in the sample) of 15.3$^{+15}_{-5}$\% for our proper motion selected subset. 
Statistical uncertainties (1~$\sigma$) throughout this paper 
were calculating using Poisson statistics for a small sample 
in the manner discussed by Burgasser et al. (2003).  

Our proper motion sample 
includes objects in the $I$ magnitude range 17.74--20.05. We note that the two primaries of 
our binaries have $I$ brighter than 18.5. We define magnitude limited bins of 
proper motion members as follows: (1) between $I$ = 17.74--18.50 we have 9 objects 
and 2 binaries; (2) between $I$ = 18.50--21.00 we have 4 objects 
and 0 binaries. We also note that 4 of the 10 Pleiades BDCs without proper motion data 
lie in the second magnitude bin. The rate of success in confirming BDs among the 
BDCs found in the CFHT survey (Bouvier et al. 1998) in the magnitude bin of subset (2) 
has been 55\%  (6 proper motion members out of 11 candidates, Moraux et al. 2001). 
If the success rate is similar for the NPL and Roque surveys, we expect 2 of the 4 BDCs 
to be confirmed. Thus, we add 2 objects to the sample in bin (2) and we get 0 binaries 
for 6 objects. 

To summarize, we get the following binary frequencies for the separation range 7-12~AU: 
F$_b$ = 2/13 = 15$^{+15}_{-5}$\% for the sample of proper motion 
members in the magnitude range $I$ = 17.74--20.05; 
F$_b$ = 2/9 = 22$^{+19}_{-8}$\% in the magnitude bin $I$ = 17.74--18.50 for proper motion members; 
and F$_b$ = 0/6 implying F$_b <$ 23\% in the magnitude bin $I$ = 18.50--21.00 
for targets with confirmed proper motion membership (4 objects) plus targets without 
proper motion information (2 out of 4 objects).  

Our survey is not sensitive to separations smaller than 7~AU in the Pleiades and smaller 
than 10.5~AU in $\alpha$Per. We do not find any companions with separations larger than 12~AU 
among the sample of 13 Pleiades proper motion members, and we do not find any binaries 
with separations larger than 10.5~AU among the sample of 8 $\alpha$Per candidate members. 
No proper motions are available for the $\alpha$Per objects, but 4 of them have lithium 
detections (Basri \& Mart\'\i n 1999; Stauffer et al. 1999). We estimate that BD 
binaries with separations larger than 12~AU are less frequent than 1/21, i.e. $<$9\% in the 
$\alpha$Per and Pleiades open clusters. This result is consistent with the low frequency 
($<$2\%) of ultracool binaries in the solar neighborhood with separations larger than 15~AU 
(Bouy et al. 2003; Close et al. 2003; Gizis et al. 2003; Reid et al. 2001).

\subsection{Comparison with theoretical models: Mass estimates} 

In Figure~4 we show a color-magnitude diagram (CMD) for the sample of proper motion 
Pleiades members 
included in our survey. The components of the 3 binaries are plotted individually. Note that 
the photometric uncertainties are higher for these objects (Table~2). 
The synthetic WFPC2 photometry from the 
Nextgen (Baraffe et al. 1998) and Dusty (Chabrier et al. 2000) isochrones for 
ages 70 and 120~Myr are shown. These ages bracket the range of cluster ages 
estimated for the Pleiades in the literature (Stauffer et al. 2003 
and references therein).  

The Nextgen isochrones give a better fit to the cluster sequence for $I<$18.5, 
and the Dusty isochrones provide a better fit for fainter objects. 
The agreement between the models and data is quite good within the error bars. 

We have estimated masses for the objects using an age of 120~Myr (LDB age, 
Stauffer \& Barrado y Navascu\'es 2003) and the Nextgen models for $I<$18.5. 
For fainter objects we used the Dusty models. The masses derived for the 
components of the 3 binaries are given in Table~2. 
CFHT-Pl-12 and IPMBD~29 have estimated orbital periods of less than 
a century. Orbital motion should be measurable in a few years. 
The dynamical masses of BD binaries of known age constitute an 
important test to the accuracy of evolutionary models.

\subsection{Binary properties as a function of primary mass} 

We have compiled all the known resolved (separation $>$ 7 AU) 
stellar binaries in the Pleiades 
from the following papers: Abt et al. (1965); Anderson, Stoeckly \&  Kraft (1966); 
Bouvier, Rigaut, \& Nadeau (1997); Jones, Fischer \& Stauffer (1996); 
Mason et al. (1993); Mermilliod et al. (1992); Raboud \& Mermilliod (1998); 
Rosvick, Mermilliod, \& Mayor (1992a,b); Stauffer (1982); and Stauffer et al. (1984). 
Their primary masses are plotted in Figure~5 with respect to their projected 
semimajor axis. There is a strong trend for the upper envelope of maximum 
separations to decrease with primary mass. A similar effect has been noted 
among binaries in the field (Burgasser et al. 2003 and references therein). 
The lack of BD binaries in the 
$\alpha$Per and Pleiades open clusters, and in the solar vicinity, with separations 
larger than 15~AU, may be a natural continuation of this effect, already present 
in stellar binaries, toward substellar masses. In order to check if there is 
any sudden change in the distribution of semimajor axis across the substellar boundary, 
a survey for binaries among M-type Pleiades members is needed. 

In section 4.1 
we obtained a binary frequency for the semimajor axis range 7--12~AU 
of F$_b$ = 22$^{+19}_{-8}$\% in the magnitude bin $I$ = 17.74--18.50, 
which corresponds to primary masses in the range 0.072 -- 0.05~M$_\odot$ according 
to the Nextgen models for an age of 120~Myr. We also derived a binary frequency 
for the same range of semimajor axis 
of F$_b <$ 23 \% in the magnitude bin $I$ = 18.50--21.00, corresponding 
to primary masses between 0.050~M$_\odot$ and 0.030~M$_\odot$ 
according to the Dusty models for an age of 120~Myr. The difference between these 
two mass bins is not statistically significant, but it suggests 
that lower mass BD binaries 
may be tighter than the resolution limit of our WFPC2 survey. 
The trend seen in Figure~5 may continue into the substellar domain. A larger sample 
of binaries is needed to improve the statistics. 

Figure~6 displays the mass ratios of resolved binaries in the Pleiades cluster, 
including the 3 BD binaries found in this survey. We note the lack of mass ratios 
equal to unity among the BD binaries. This may be an effect of low number statistics.  
We also note that there are no BD binaries with q$<$0.7. The sensitivity of our 
survey is $I$ = 22.5, which translates into a mass of 0.027~M$_\odot$ using 
the Dusty models for an age of 120~Myr. Thus, we are sensitive to mass ratios 
q$>$0.37 for our most massive primaries (Mp$\sim$0.07~M$_\odot$), and larger q for 
lower mass primaries. However, all our BD binaries were found 
at close separations where the sensitivity limit is only about 3 magnitudes 
(Bouy et al. 2003). The lack of BD binaries in our survey with mass ratios 
in the range q=0.7--0.4 may be an observational effect. Since our survey 
is not as sensitive to low mass ratios as other surveys (e.g., Bouvier et al. 1997), 
our BD binary frequencies may be underestimated if there is a significant 
number of BD binaries with mass ratios q$<$0.7 which we have missed. 
Nevertheless, we note that high-resolution surveys of nearby ultracool dwarfs 
have also failed to find binaries with q$<$0.7 in larger samples (Bouy et al. 2003; 
Close et al. 2003; Gizis et al. 2003). So, there could be a real trend 
for VLM binaries to have mass ratios larger than q=0.7. 

In Figure~7 we plot the frequency of binaries wider than 7~AU as a function of 
primary mass  in the Pleiades cluster. The binary frequency for brown dwarfs  
does not appear to be lower than for stars, although 
the error bars are still high due to low number statistics.

\section{Final remarks and future prospects} 

In a survey carried out with HST/WFPC2, 
we have found 4 binaries in a sample of 25 candidate VLM members in the Pleiades cluster 
and no binaries in a sample of 8 candidate VLM members in the $\alpha$Persei cluster. 
We argue that 3 of the Pleiades binaries are cluster members and one is not on the basis 
of the available proper motion and spectroscopic information. 
Component masses and orbital periods are estimated from theoretical evolutionary models. 
Follow-up observations will yield orbital parameters in a timescale of about a decade, 
and will provide powerful tests to the theory of substellar-mass objects. 

No binaries are detected with separations larger than 12~AU among the VLM population 
of the $\alpha$Persei and Pleiades clusters. We estimate a frequency of $<$9\% for 
VLM binaries with separations larger than 12~AU in these clusters. This result 
agrees with the scarcity of VLM binaries wider than 15 AU in the solar neighborhood 
($<$2\%,  Bouy et al. 2003; Close et al. 2003; Gizis et al. 2003). 
We show that in the Pleiades there 
is a decrease in the upper envelope of binary separations with decreasing primary mass. 
The VLM binaries may be a natural continuation of this effect. We note that there 
is a significant lack of data of binaries for masses in the range 0.3-0.1~M$_\odot$ 
in the Pleiades. A study of binarity in those stars is needed to fill the gap 
between the cool stars and the VLM objects. 

We find that the binary frequency among Pleiades BDs with confirmed proper motion 
membership in the mass range 
0.072 -- 0.030~M$_\odot$ and separations larger than 7~AU is 15\% . It is similar to that 
the binary fraction of stellar binaries for the same separation range. 
We divide our Pleiades sample in two mass bins. 
We estimate a binary frequency of 22$^{+19}_{-8}$\%  for primary masses in the range 
0.072 -- 0.050~M$_\odot$. For primary masses 
between 0.050~M$_\odot$ and 0.030~M$_\odot$ we derive an upper limit 
of $<$23\%. Although our study is hampered by low number statistics, 
we suggest that low mass BD binaries 
may be systematically tighter than high-mass BDs, and thus are not detected in our 
HST survey. Studies of larger samples of BDs are needed to test our suggestion. 

The BD binary frequency appears to be larger than 10\% in the Pleiades 
cluster and in the solar neighborhood. Such a relatively high binary frequency 
does not support hydrodynamical models of BD formation that predict a very low 
($<$5\%) BD binary frequency (Bate et al. 2002). Within the error bars, our results may be 
consistent with dynamical decay models of evolution in open clusters with random pairing  
(Sterzik \& Durisen 2003). However, such models predict a flat mass ratio distribution 
which does not seem to agree with the observed lack of VLM binaries with q$<$0.7. 
Models of BD formation should take into account the fundamental observational constrains 
that BD binaries are numerous, and that they are biased toward separations 
smaller than 15~AU and that the mass ratios tend to be larger than 0.7. 

Interferometric imaging and 
radial velocity studies of the Pleiades VLM population are needed to discover    
short period binaries such as PPl~15 (Basri \& Mart\'\i n 1999) and determine 
their frequency and mass ratio distribution. Larger samples of BDs in clusters 
should be studied to improve the statistical significance of the BD binary properties.

\acknowledgments

Support for Proposal number 8701 was provided by NASA through a grant from the 
Space Telescope Science Institute, which is operated by the Association of Universities 
for Research in Astronomy, Incorporated, under NASA contract NAS5-26555.
Support for this work was provided by National Aeronautics and Space
Administration(NASA) grant NAG5-9992 and National Science Foundation(NSF)
grant AST-0205862. 
This paper is partly based on data collected 
the Keck~I telescope at the W. M. Keck Observatory, Mauna Kea, Hawaii. The Keck observatory is 
operated as a scientific partnership among the California Institute of Technology, 
the University of California, and the National Aeronautics and Space Administration. 
The Observatory was made possible by the generous financial support of the W. M. Keck 
Foundation.  
This research has made use of the SIMBAD database, operated at CDS, Strasbourg, France. 
The authors wish to extend special thanks to those of Hawaiian ancestry on whose sacred 
mountain of Mauna Kea we are privileged to be guests. Without their generous hospitality, 
the Keck telescope observations presented therein would not have been possible.  
We thank Leif Festin, Richard Jameson, John Stauffer and Maria Rosa Zapatero Osorio for 
helping with the initial selection of targets for HST observations. 
John Gizis provided a helpful referee report 
that contributed to improve the paper significantly.

\clearpage


\begin{figure}[h]
\plotone{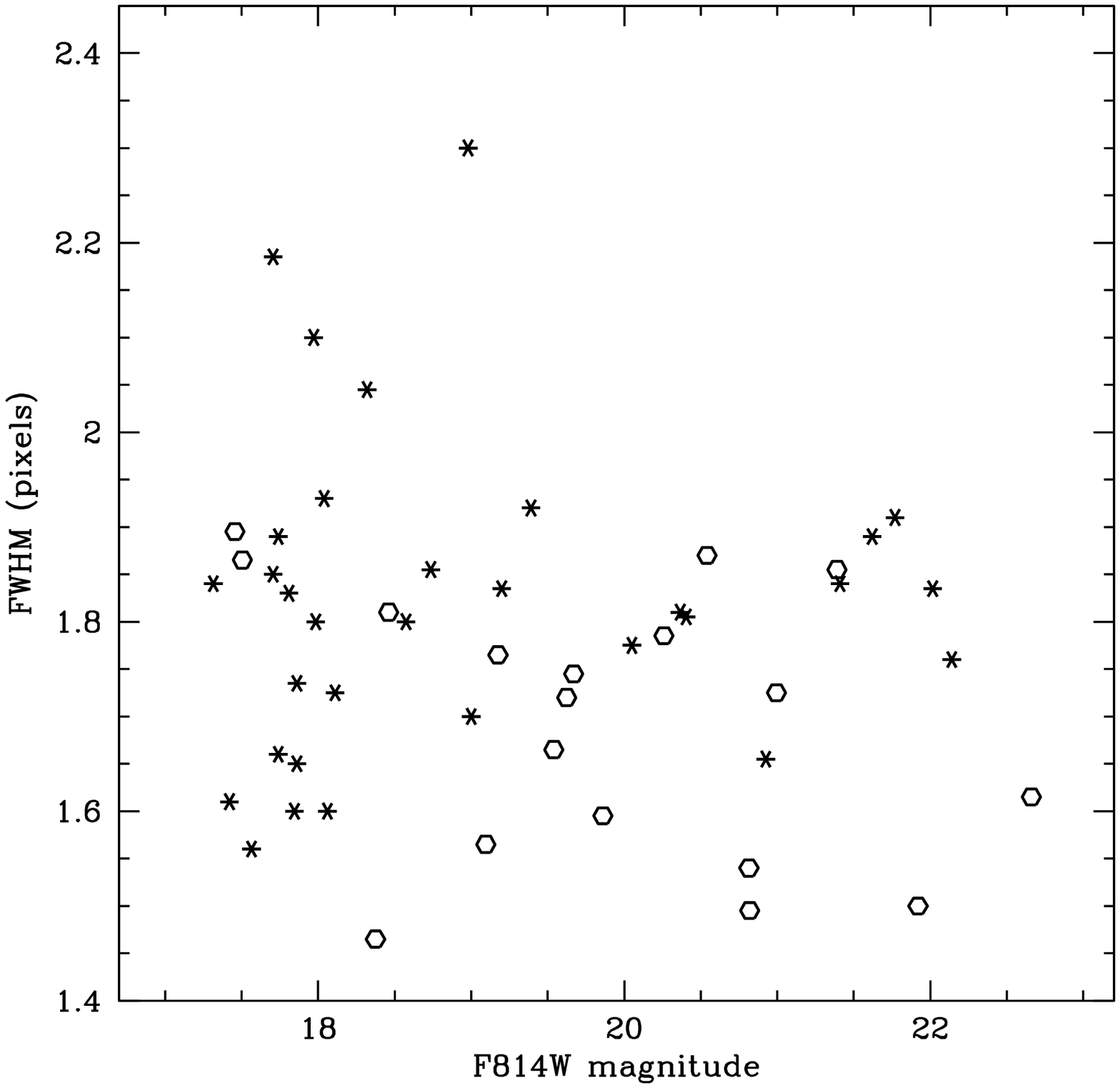}
\caption{Full width half maximum (FWHM) measurements in the F814W filter for our program objects 
(asterisks) 
and for reference stars in the PC1 field of view (empty hexagons). The four objects with FWHM$>$2 
are binaries.}
\end{figure}

\begin{figure}[h]
\vbox to4in{\rule{0pt}{4in}}
\includegraphics{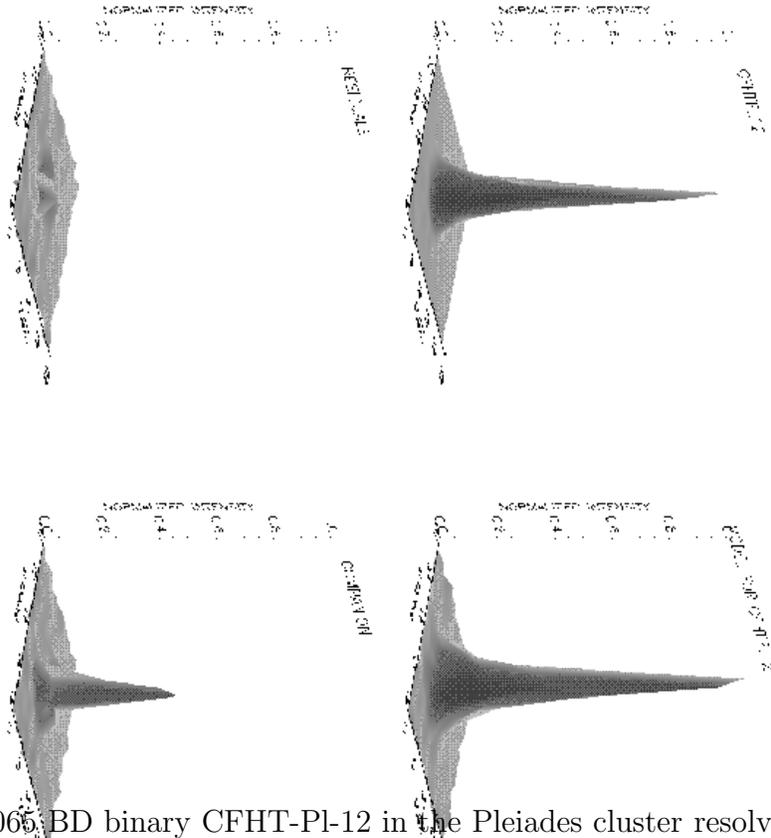}
\caption{The 0$\arcsec$.065 BD binary CFHT-Pl-12 in the Pleiades cluster resolved with 
our PSF analysis of WFPC2 F814W observations. 
Upper left: Binary PSF. Upper right: Model PSF. 
Lower right: Residual after subtraction of model PSF for primary object. 
Lower left: Residual after subtraction of model PSF for secondary object.}
\end{figure}

\begin{figure}
\plotone{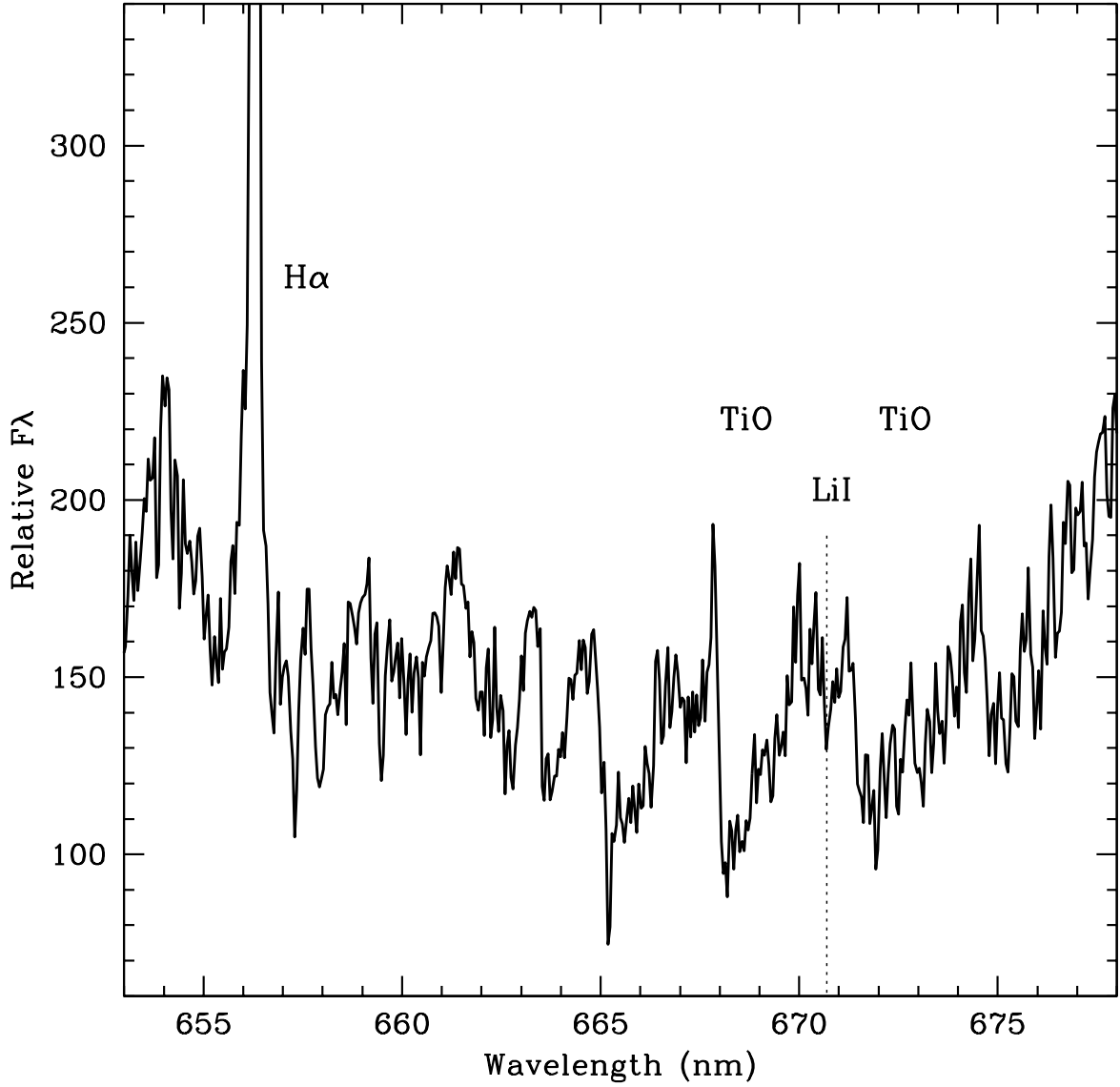}
\caption{A portion of the ESI spectrum of IPMBD~25 showing the detection of H$_\alpha$ in 
emission and LiI absorption.}
\end{figure}

\begin{figure}
\plotone{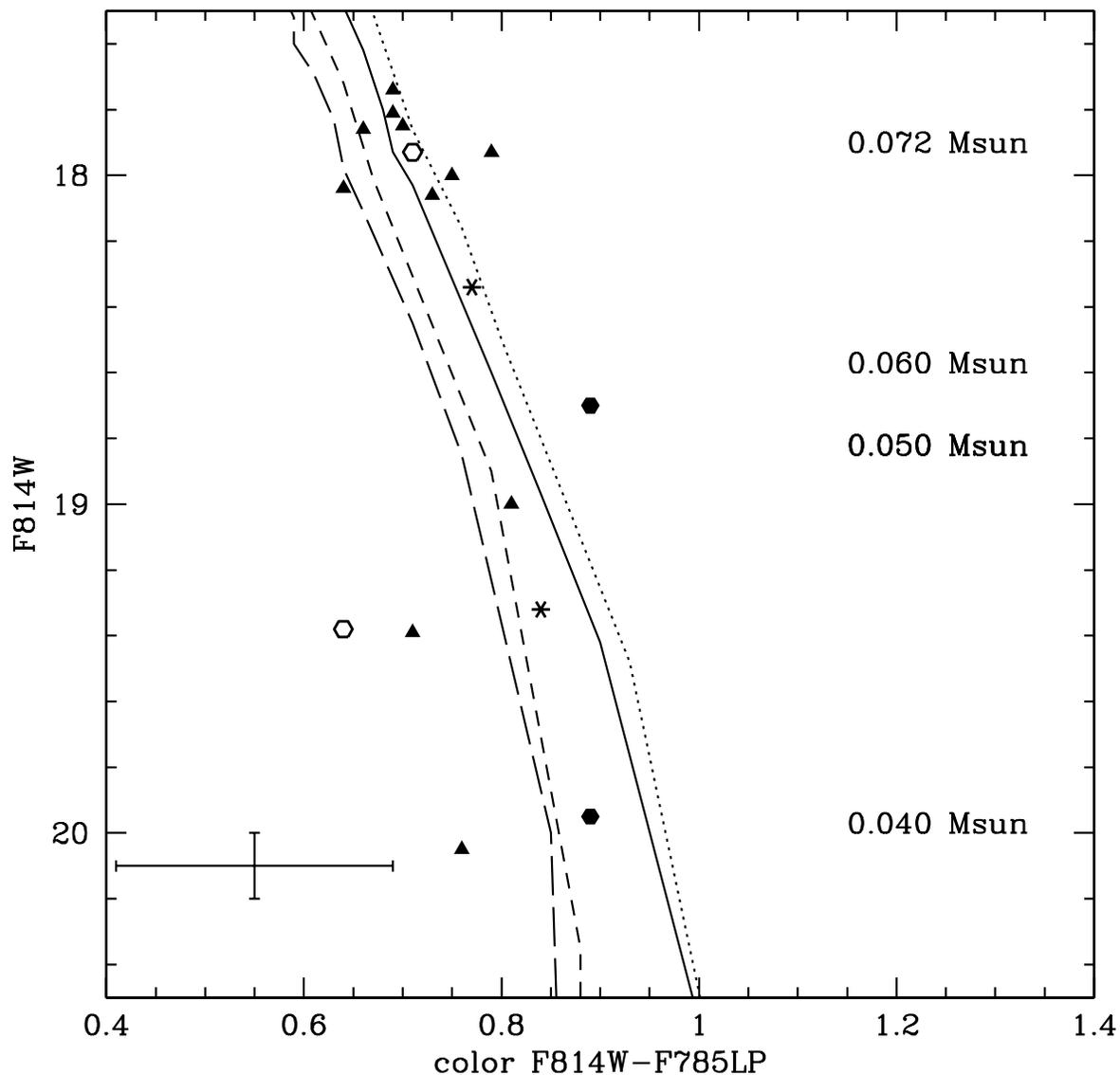}
\caption{Color-magnitude diagram for proper motion Pleiades members included in our WFPC2 survey. 
The components of 3 resolved binaries share the same symbols: CFHT-Pl-12 (asterisks), 
IPMBD~25 (open hexagons) and IPMBD~29 (filled hexagons). Unresolved targets are denoted with 
filled triangles. The following theoretical isochrones are overplotted from right to left: 
Nextgen--70~Myr 
(dotted line), Nextgen--120~Myr (solid line), Dusty--70~Myr (short dashed line), and Dusty--120~Myr 
(long dashed line). Masses obtained from the Nextgen--120~Myr (0.072 and 0.060~M$_\odot$) and the 
Dusty--120~Myr (0.050 and 0.040~M$_\odot$) are labelled. A typical errorbar for the photometric data 
is shown in the lower left corner.}
\end{figure}

\begin{figure}
\plotone{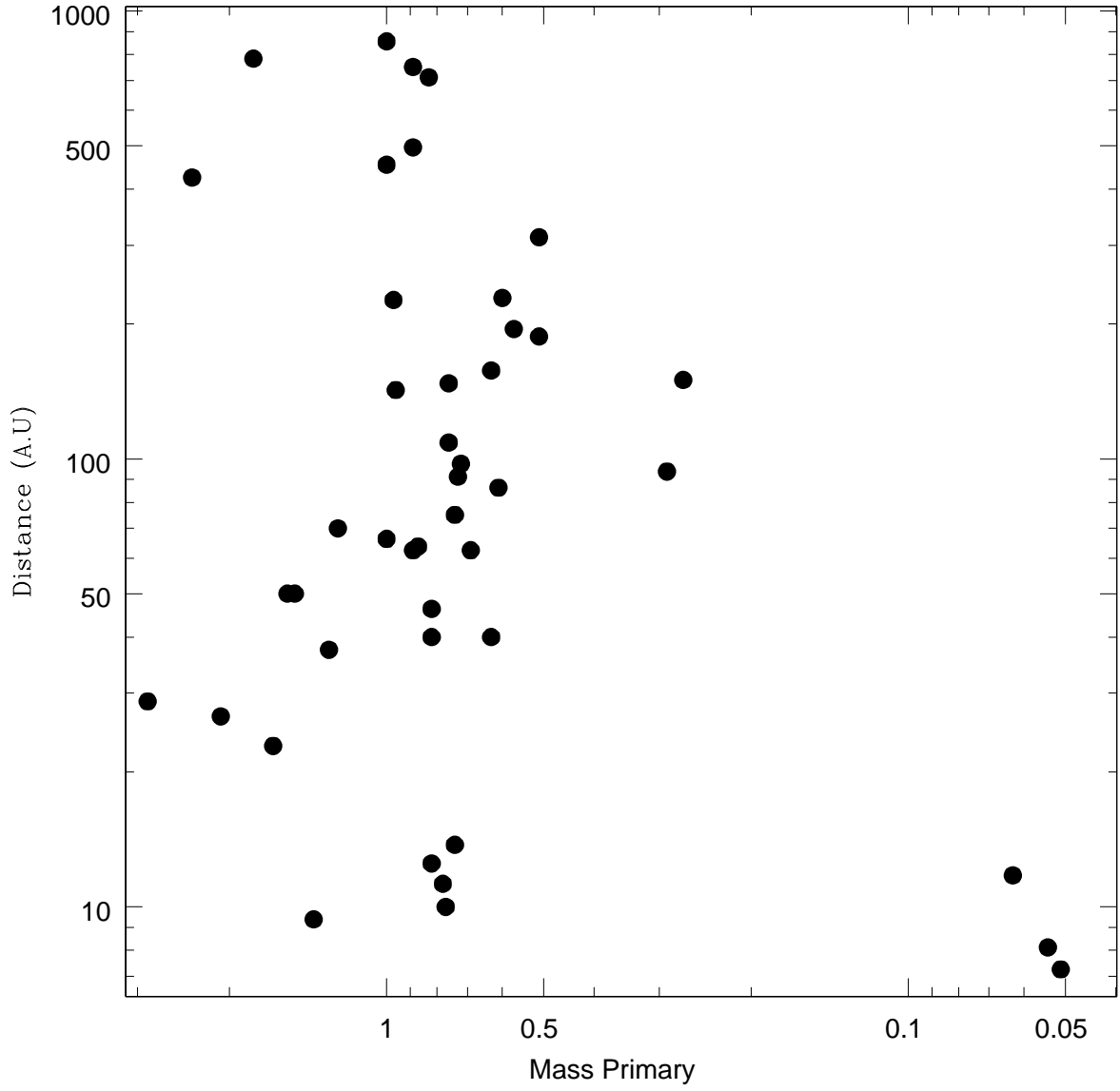}
\caption{Primary mass versus semimajor axis for resolved (separation $>$ 7 AU) 
binaries in the Pleiades cluster.}
\end{figure}

\begin{figure}
\plotone{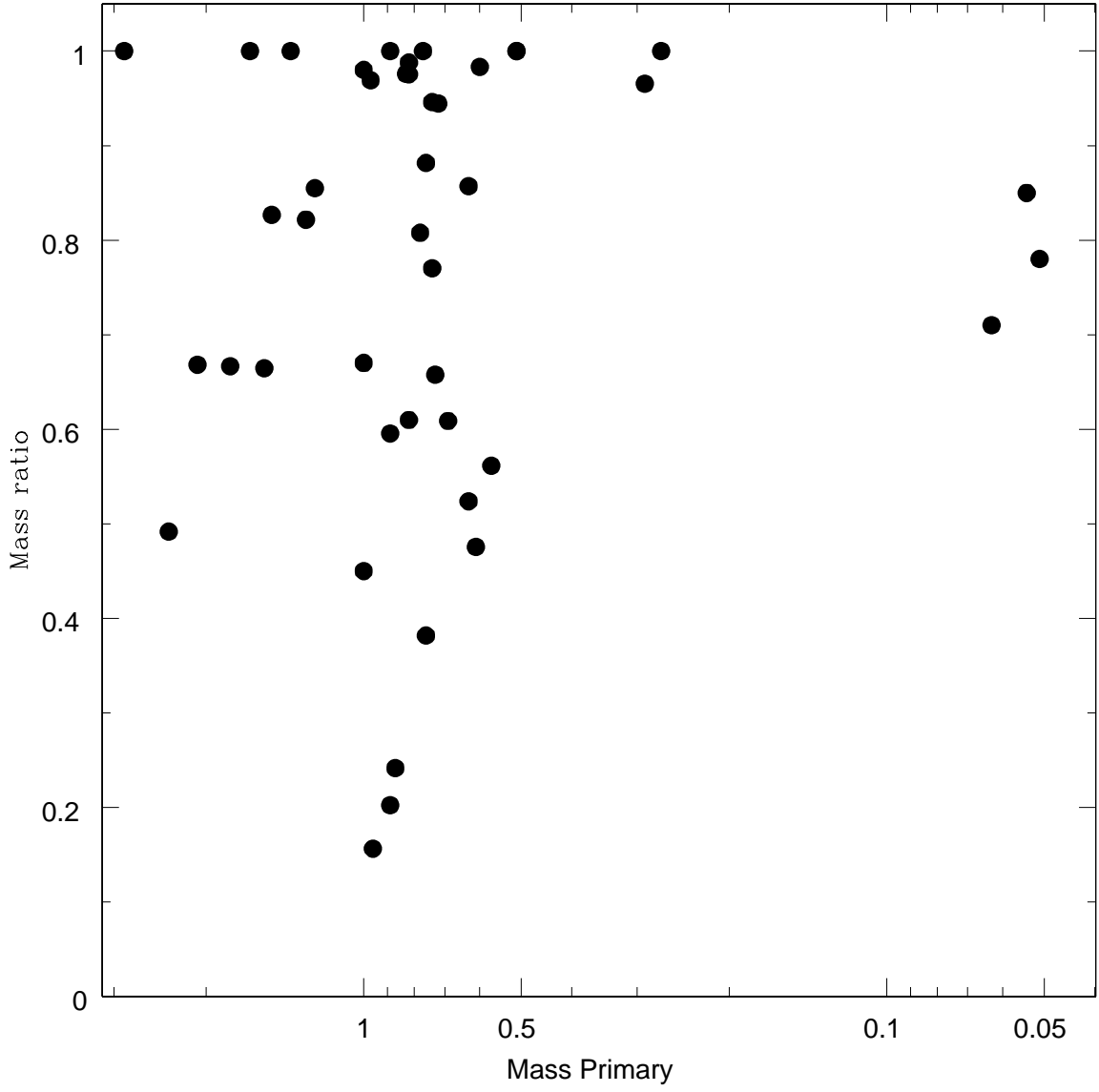}
\caption{Primary mass versus mass ratio for resolved binaries in the Pleiades cluster.}
\end{figure}

\begin{figure}
\plotone{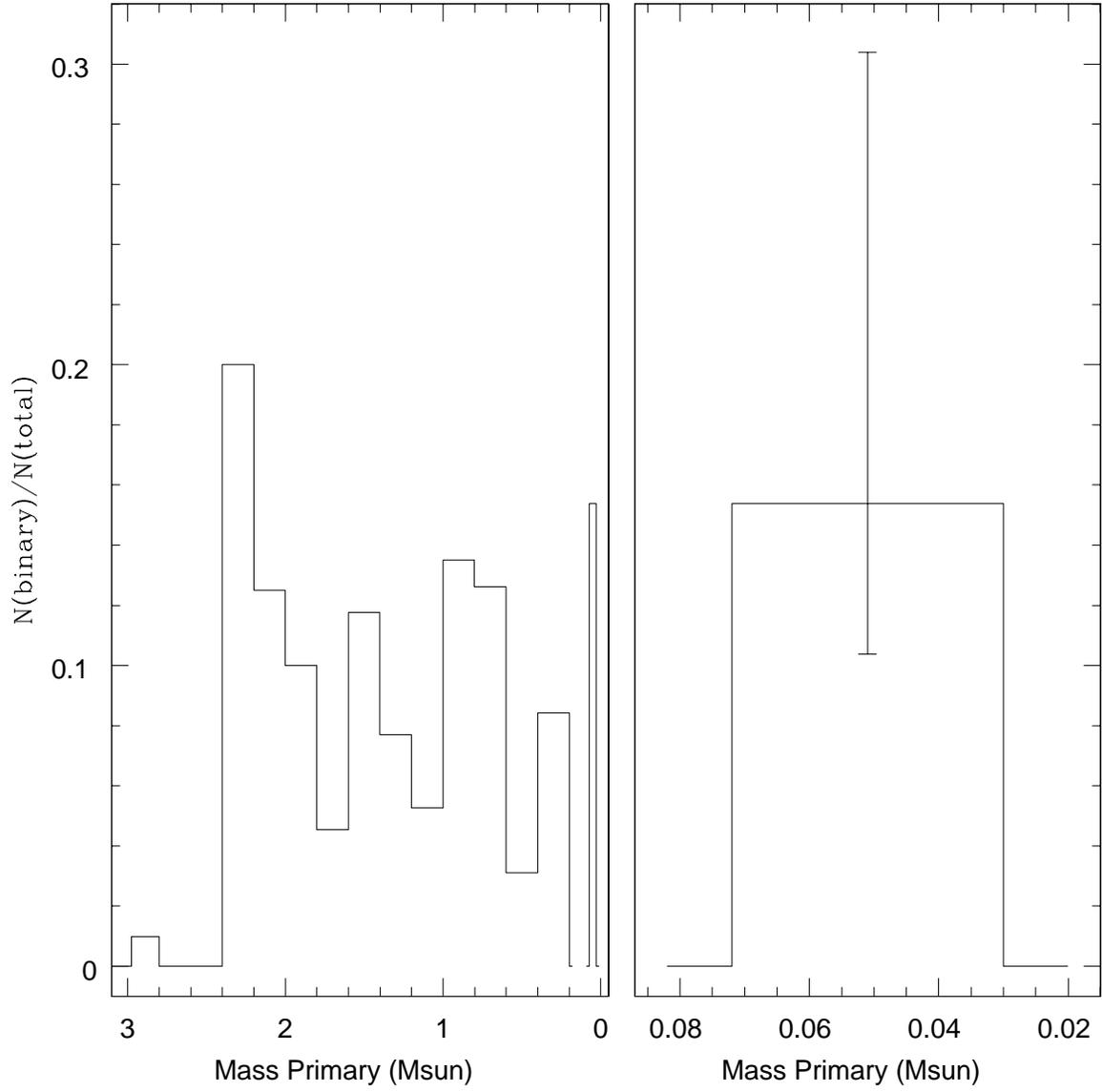}
\caption{Frequency of binaries wider than 7~AU as a function of 
primary mass  in the Pleiades cluster. The right panel is a zoom on the 
substellar-mass regime of this diagram.}
\end{figure}

\clearpage

\begin{deluxetable}{llllllllll}
\tabletypesize{\scriptsize}
\tablecaption{Data for program objects}
\tablewidth{0pt}
\tablehead{
\colhead{Name} & 
\colhead{F814W} & 
\colhead{FWHM} &
\colhead{F785LP} & 
\colhead{FWHM} &
\colhead{$I_{\rm C}$\tablenotemark{a}} &
\colhead{$R-I_{\rm C}$} &
\colhead{SpT} &
\colhead{Li} &
\colhead{PMM}  
}
\startdata
Calar~3     & 19.00 & 1.70 & 18.19 & 1.83 & 19.00 & 2.50 & M8     & Y   & Y  \\
CFHT-Pl-9   & 17.74 & 1.66 & 17.05 & 1.81 & 17.71 & 2.18 & M6.5   & N   & Y  \\
CFHT-Pl-10  & 17.81 & 1.83 & 17.12 & 1.97 & 17.82 & 2.21 & M6.5   & N   & Y  \\
CFHT-Pl-12  & 17.93 & 2.10 & 17.14 & 2.15 & 18.00 & 2.47 & M8     & Y   & Y  \\
CFHT-Pl-15  & 18.73 & 1.85 & 17.96 & 1.85 & 18.62 & 2.34 & M7     & Y   & N  \\
CFHT-Pl-19  & 18.99 & 2.30 & 18.30 & 2.27 & 18.92 & 2.51 & na\tablenotemark{b} & na  & N  \\
IPMBD~11    & 18.00 & 1.80 & 17.25 & 1.99 & 18.07 & na   & na     & na  & Y  \\
IPMBD~21    & 17.85 & 1.60 & 17.15 & 1.80 & 17.85 & na   & na     & na  & Y  \\
IPMBD~22    & 17.86 & 1.65 & 17.20 & 1.86 & 17.90 & na   & na     & na  & Y  \\
IPMBD~25    & 17.67 & 2.19 & 16.97 & 2.12 & 17.82 & na   & na     & na  & Y  \\
IPMBD~26    & 18.04 & 1.93 & 17.40 & 1.79 & 18.11 & na   & na     & na  & Y  \\ 
IPMBD~29    & 18.38 & 2.05 & 17.49 & 2.10 & 18.35 & na   & na     & na  & Y  \\
IPMBD~43    & 18.06 & 1.60 & 17.33 & 1.73 & 18.1: & na   & na     & na  & Y  \\
NPL~36      & 18.56 & 1.80 & 17.79 & 2.00 & 18.66 & na   & M7.5   & na  & na \\
NPL~38      & 19.20 & 1.83 & 18.38 & 1.84 & 19.18 & na   & M8     & na  & na \\
NPL~43      & 21.77 & 1.91 & 21.05 & 1.82 & 21.79 & na   & na     & na  & na \\
Roque~5     & 20.05 & 1.78 & 19.29 & 1.72 & 19.71 & na   & M9     & na  & Y  \\ 
Roque~7     & 19.39 & 1.92 & 18.68 & 1.85 & 19.50 & 2.61 & M8.5   & na  & Y  \\
Roque~18    & 21.41 & 1.84 & 20.85 & 1.81 & 21.11 & na   & na     & na  & na \\
Roque~20    & 21.62 & 1.89 & 22.31 & 1.59 & 22.2: & na   & na     & na  & na \\
Roque~23    & 22.01 & 1.84 & 21.73 & 1.83 & 21.75 & na   & na     & na  & na \\
Roque~24    & 22.14 & 1.76 & 21.02 & 1.79 & 21.56 & na   & na     & na  & na \\
Roque~30    & 20.92 & 1.67 & 20.40 & 1.86 & 20.31 & na   & na     & na  & na \\
Roque~33    & 20.39 & 1.81 & 19.65 & 1.75 & 20.26 & na   & M9.5   & na  & na \\   
AP~270      & 17.80 & 1.80 & 17.01 & 1.75 & 17.83 & na   & M6     & Y   & na \\
AP~275      & 17.31 & 1.84 & 16.62 & 1.66 & 17.25 & 2.20 & M6     & N   & na \\ 
AP~300      & 17.86 & 1.73 & 17.17 & 1.93 & 17.85 & 2.18 & M6     & Y   & na \\
AP~310      & 17.71 & 1.85 & 17.08 & 1.71 & 17.80 & 2.33 & M6     & N   & na \\
AP~318      & 17.42 & 1.61 & 16.75 & 1.64 & 17.45 & 2.16 & M6     & N   & na \\
AP~323      & 17.57 & 1.56 & 16.92 & 1.77 & 17.50 & 2.13 & na     & Y   & na \\
AP~324      & 18.11 & 1.72 & 17.44 & 1.63 & 18.10 & 2.36 & M6.5   & Y   & na \\
AP~325      & 17.74 & 1.89 & 16.94 & 1.78 & 17.65 & 2.30 & M7     & Y   & na \\
\enddata
\tablenotetext{b}{The $R_{\rm C}$ and $I_{\rm C}$ photometric data, spectral types, 
lithium detections and proper motion memberships come from the literature 
cited in Section 2.}
\tablenotetext{b}{na=not available.}
\end{deluxetable}

\begin{deluxetable}{lllllllllll}
\tabletypesize{\scriptsize}
\tablecaption{Data for Binary Systems.}
\tablewidth{0pt}
\tablehead{
\colhead{Name} & 
\colhead{F814W(A)} & 
\colhead{F814W(B)} &
\colhead{F785LP(A)} &
\colhead{F785LP(B)} & 
\colhead{Sep (arcsec)} & 
\colhead{Sep (AU)} &
\colhead{P.A.} &
\colhead{M$_{\rm p}$\tablenotemark{a}} & 
\colhead{q} &
\colhead{P(yr)\tablenotemark{b}} 
 }
\startdata
CFHT-Pl-12 & 18.34$\pm$0.11 & 19.32$\pm$0.11 & 17.57$\pm$0.11 & 18.48$\pm$0.11 & 0.062$\pm$0.002 
& 7.75  & 266.7$\pm$1.7 & 0.054 &  0.70 &  76   \\
CFHT-Pl-19 & 19.40$\pm$0.11 & 20.38$\pm$0.11 & 18.79$\pm$0.11 & 19.57$\pm$0.11 & 0.066$\pm$0.003 
&       & 262.7$\pm$1.8 &       &       &         \\
IPMBD~25   & 17.93$\pm$0.09 & 19.38$\pm$0.09 & 17.22$\pm$0.09 & 18.74$\pm$0.09 & 0.094$\pm$0.003 
& 11.75 & 340.5$\pm$2.1 & 0.063 &  0.62 & 126   \\
IPMBD~29   & 18.70$\pm$0.15 & 19.95$\pm$0.15 & 17.81$\pm$0.11 & 19.06$\pm$0.11 & 0.058$\pm$0.004 
& 7.25  & 103.0$\pm$4.5 & 0.045 &  0.84 &  68   \\
\enddata
\tablenotetext{a}{Masses for primaries (M$_{\rm p}$) are given in solar masses 
(see text for details).}
\tablenotetext{b}{
Orbital periods are estimated for circular orbits using Kepler's third law and are given in years. 
No masses and periods 
are estimated for CFHT-Pl-19 because it is thought to be a nonmember.}
\end{deluxetable}

\end{document}